# Photonic nodal lines with quadrupole Berry curvature distribution

Dongyang Wang[1], Biao Yang[1,2], Ruo-Yang Zhang[1], Wen-Jie Chen[3], Z. Q. Zhang[1], Shuang Zhang[4,5]*, C. T. Chan[1]*

**Affiliations:**
[1]Department of Physics, Hong Kong University of Science and Technology, Hong Kong, China.
[2]College of Advanced Interdisciplinary Studies, National University of Defense Technology, Changsha 410073, China.
[3]School of Physics & State Key Laboratory of Optoelectronic Materials and Technologies, Sun Yat-Sen University, Guangzhou 510275, China.
[4]Department of Physics, The University of Hong Kong, Hong Kong, China.
[5]Department of Electrical & Electronic Engineering, The University of Hong Kong, Hong Kong, China.

*Correspondence to: shuzhang@hku.hk; phchan@ust.hk

**Abstract:**

In periodic systems, nodal lines are loops in the three-dimensional momentum space where two bands are degenerate with each other. Nodal lines exhibit rich topological features as they can take various configurations such as rings, links, chains and knots. These line nodes are usually protected by mirror or *PT* symmetry. Here we propose and demonstrate a novel type of photonic straight nodal lines in a $D_{2d}$ meta-crystal which are protected by roto-inversion time (roto-PT) symmetry. The nodal lines are located at the central axis and hinges of the Brillouin zone. They appear as quadrupole sources of Berry curvature flux and allow for the precise control of the quadrupole strength. Interestingly, there exist topological surface states at all three cutting surfaces, as guaranteed by the π-quantized Zak phases along all three directions. As frequency changes, the surface state equi-frequency contours evolve from closed to open contours, and become straight lines at a critical transition frequency, at which diffraction-less surface wave propagation are demonstrated, paving way towards development of super-imaging photonic devices.

Topological band theories provide photonic systems with driving forces for both exploration of new fundamental physics, and new insights for designing novel functioning devices(*1-4*). Topological phases have been widely demonstrated in photonic systems, including quantum Hall photonic crystals(*5-8*), photonic topological insulators(*9-11*), valley photonics(*12-14*), Weyl/Dirac semimetals(*15-20*), high-order topology(*21-26*) and non-Abelian frame charges(*27-32*). Practical photonic components based on the concept of topology have also been proposed, examples include back scattering immune waveguides(*7*), robust optical delay lines(*33*), and topological lasers(*34-37*).

Photonic semimetals have recently attracted much attention due to the singular crossings of bulk bands, where intriguing topological boundary correspondences are expected. More interestingly, 3D semimetals exhibiting line nodes of various configurations in the momentum space have been discovered, such as nodal rings, nodal chains, nodal links(*38-48*), and more recently, straight nodal lines(*49, 50*). Nodal lines were also protected by non-Abelian frame rotation charges(*27, 31, 32*). These exotic photonic line nodes are protected either by *PT* symmetry or mirror symmetry, and are usually accompanied with drumhead surface states protected by π-Zak phase(*51*). By breaking the protecting symmetries, the nodal lines could transit to Weyl points or insulating phases with toroidal moment in momentum space(*52*).

In this paper, we present a novel type of line nodes that are protected by the combined roto-inversion symmetry and time reversal *T* symmetry, dubbed as, roto-*PT* symmetry. The roto-*PT* symmetry pins the straight nodal lines to Brillouin zone (BZ) center and hinges. These nodal lines possess quadratic dispersion in contrary to the commonly studied ones with linear dispersion. They split into linear dispersion pairs when the underlying symmetries are broken. These nodal lines traverse the whole BZ and as a result, the associated drumhead surface modes can spread across the whole surface BZ. More interestingly, the hinge-location of straight nodal lines allows the equ-frequency contours (EFCs) of surface modes to transform from closed to open (so as to connect to the hinge nodal lines projections) as frequency changes. The open EFCs (i.e., photonic "Fermi-arcs") become straight lines traversing the whole surface BZ at a transitional frequency, which exhibit uniform group velocity in momentum space, allowing for diffractionless beam propagation that carries sub-wavelength information, leading to the super-imaging effect. Even though the top and bottom surfaces have different exposed structures, each "Fermi-arc" on one surface has a counterpart on the opposite side of the sample due to roto-inversion symmetry. These "Fermi-arc" pairs on opposite surfaces intersect at positions in the BZ corresponding to quantized Zak phase of π.

Here we demonstrate that roto-*PT* ($S_4T$) symmetry protected straight nodal lines can exist a photonic meta-crystal with $D_{2d}$ point group symmetry. The symmetry operation $S_4T$ converts ($k_x$, $k_y$, $k_z$) to ($k_y$, -$k_x$, $k_z$), which ensures degeneracy when $|k_x| = |k_y| = 0$ or π, i.e. at the BZ center and hinges (Supplementary Material section 1). These straight nodal lines can be viewed as the origins of Berry curvature, where quadrupole distribution of in-plane Berry curvature are ensured by the $D_{2d}$ point group together with the time reversal symmetry (Supplementary Material section 2, Fig. S1). The designed meta-crystal structure is shown in Fig. 1A, wherein the two metallic bars are of the same size as $L_1 = L_2 = 3$ mm, the length of cross bar is $L_c = 3.5$ mm, and the vertical cylinder is of $L_z = 1.5$ mm height. There are two mirror symmetries $M_{x,y}$, one roto-inversion symmetry $S_{4z}$ and two $C_2$ classes ($C_{2z}$ along z-axis and $C_{2\pm}$ along x = ± y directions), which form elements of the $D_{2d}$ point group. Band structures for the designed meta-crystal are calculated using the commercial software CST microwave studio and shown in Fig. 1B. It shows that the 1$^{st}$ and the 2$^{nd}$ bands come into degeneracy along the Γ – Z line (blue), while the 2$^{nd}$ and the 3$^{rd}$ bands are degenerate along the M – A line (red). We show the degeneracy line nodes in the BZ in Fig. 1C (left panel), where the straight nodal lines are displayed as red and blue lines located at the two high symmetry lines of the BZ cuboid (through Γ – Z and M – A).

The photonic straight nodal line systems are accompanied by topological surface modes, which can be analyzed with the Zak phase $\phi$ integrated along multiple directions. For the proposed model, the two mirror symmetries of $M_{x,y}$ quantize the Zak phases $\phi_{x,y}$ along x and y directions for the whole mirror planes, as indicated in Fig. 1C (right panel) with red and blue arrows and planes. In addition, the $C_{2\pm}T$ symmetry quantize the Zak phase $\phi_z$ for the loops along the z-direction (e.g. green arrow embedded in diagonal plane). These quantized Zak phases lead to three $Z_2$ indices that characterize the surface modes with different surface cut. Particularly, for each point on the path of Z – A, we can numerically calculate the Zak phase along all three directions (Supplementary Material section 3 and Fig. S2), and show the total contribution to the electromagnetic wave macroscopic polarization (i.e. surface mode) in Fig. 1D for the gap between the 2$^{nd}$ and 3$^{rd}$ bands (summation of the quantized Zak phase of the 1$^{st}$ and 2$^{nd}$ bands)(*53*). Interestingly, all three Zak phases are quantized to π. The combined indices of [$Z_{2x}$, $Z_{2y}$, $Z_{2z}$] = [1, 1, 1] thus guarantee the presence of topological surface modes on the interfaces truncated normal to all three orthogonal directions. In a similar manner to the common drumhead surface states in nodal line systems, the surface modes here connect to the projection of straight nodal lines as well, which promises the shape transition of the EFCs of the surface mode with frequency from closed to open contours and allows for the surface wave manipulation demonstrated in experiments shown below.

Although there are two mirror symmetries in the D$_{2d}$ group, it should be pointed out that the straight nodal lines are roto-*PT* symmetry protected (more discussion in Supplementary Material section 4, Fig. S3). The mirror symmetries only protect the nodal lines existing in the mirror planes, but do not pin them to the BZ center and hinges. The roto-*PT* symmetry endows the straight nodal line with quadratic dispersions and a loop encircling the quadratic nodal line gives 0/2π Berry phase instead of π as indicated in Fig. 1C. When the $S_{4z}$ symmetry is broken, the point group reduces from D$_{2d}$ to C$_{2v}$, and each of the quadratic nodal lines splits into two linear ones. We show the transformation of band dispersion in Fig. 1E, where the quadratic touching (left column) splits into linear pairs (right column). In Fig. 1F, we show the line-nodes, each carrying a π Berry phase if one integrates the Berry connection around a tight encircling loop. More interestingly, the center straight nodal line transforms into an in-plane nodal chain, as a result of the intrinsic symmetry of photonic systems at Γ point(*32*). Besides the quadratic nodal lines, four Weyl points can be found along the *Γ-M(M')* directions between the two higher bands (the 3$^{rd}$ and 4$^{th}$), which are induced by the broken inversion symmetry.

To experimentally realize and characterize the straight nodal line system, meta-crystal samples are fabricated with printed circuit boards (PCBs) technique, where a substrate material (F4B, $\varepsilon \approx 2$) is chosen to support the 3D metallic resonators. Surface fields are experimentally scanned, and subsequently Fourier transformed to retrieve the momentum space information.

For a sample with layers stacked along the (001) direction, the top surface configuration is shown in Fig. 2A. The corresponding surface BZ is shown in Fig. 2B, where the straight nodal lines project into singular points at the center and corners of the BZ. The corresponding projection band dispersions are shown in Fig. 2C (left panel). The nodal line along $\Gamma - Z$ is manifested as a series of band pairs that become degenerate at the $\Gamma$ point. By implementing a perfect electric conductor (PEC) boundary on the top surface, we experimentally measured the band projections, which are shown in Fig. 2C (right panel). The band pairs are found to be degenerate at $\Gamma$ point as expected, which experimentally verify the center nodal line along $\Gamma - Z$ in the BZ. In a different configuration of the (100) surface, we also experimentally measured the straight nodal lines located at the BZ hinges, the results are shown in Supplementary Material section 5, and Fig. S4.

The band projections with air boundaries are calculated in Fig. 2D (left panel), which show that surface modes in the projected band gap connect from the air cone to the $\overline{M}$ point (the projection of red nodal line), similar to drumhead surface states in nodal line systems. We experimentally measured these surface modes as shown in Fig. 2D (right panel), where good agreement with the calculation can be found. In Fig. 2E, the EFCs of surface modes on the top surface for a few frequencies are shown. At a lower frequency of $f = 13.3$ GHz, the EFC is closed and surrounds the projected bulk bands. However, as frequency increases and approaches the degeneracy energy of straight nodal line along $M - A$, the EFCs of surface modes connect to the bulk degeneracy position of surface BZ corner as shown in Fig. 2E at $f = 16.9$ GHz (and also predicted in Fig. 2D). The connection to surface BZ corner forces the closed EFC to open up, the EFCs thus evolve from closed to open curves with the increase of frequency as shown in Fig. 2E.

The evolution of "Fermi-arcs" provides the topological meta-crystal with the ability to tailor the wavefront of surface wave. The surface wave propagation is experimentally investigated by using the setup shown schematically in Fig. 3A, where a point source (red color) is positioned at the corner for surface mode excitation, and another probe (blue color) is positioned close to the surface for raster scans of the field pattern. At $f = 15.9$ GHz, the surface wave is found to propagate towards the y-direction along the sample edge, as shown with the measured field in Fig. 3B. The collimated surface beam is a result of the nearly constant group velocity of the surface modes with highly flat "Fermi-arc" shown in Fig. 3C.

Flat EFCs carrying large wave vectors are desirable for the realization of sub-wavelength imaging(*54*). In Fig. 3D, a source antenna is attached to two metal stripes at the edge of the sample top surface. They function as two sources for exciting surface waves (detailed setup is shown in Supplementary Material section 6, Fig. S5). The metal stripes are 2 mm wide and separated by 5 mm as indicated in Fig. 3D. Measured propagation field and absolute amplitude are shown for $f = 15.9$ GHz in Figs. 3E and F, respectively. As can be seen, two beams propagate along y-direction with well-maintained beam profiles. The distance between the two beam centers is measured to be $d \approx 7.5$ mm and the surface wavelength is measured as $\lambda \approx 19$ mm. Thus, super-imaging is experimentally demonstrated for the topological "Fermi-arcs". It is worth noting that the EFCs at all three surfaces share similar features of shape evolution, since the hinge straight nodal line will always project to surface BZ corners. The surface modes on the three surfaces will all support super-imaging.

We note that the topological protection of "Fermi-arcs" due to quantized Zak phase is a bulk property, which does not distinguish between the two opposite surfaces of sample with symmetry-preserved truncation. Taken the (001) surface as example, the surface modes should exist at both top and bottom surfaces at a given frequency, and the $S_{4z}$ symmetry rotates the two EFCs from opposite surfaces by 90° in the $k_x$-$k_y$ plane. These two EFCs would thus form into intersection when opposite surfaces meet. Particularly, the location of intersection corresponds to the exact position of π-quantized Zak phase. We further show the experimental verification with a waveguide-like configuration in supplementary material section 7, Fig. S6, where two opposite surfaces are measured together, and the "Fermi-arc" degeneracies are observed.

The straight nodal lines are both sources and sinks of Berry curvature, and lead to quadrupolar distributions of in-plane Berry curvature in momentum space. In order to characterize the Berry curvature quadrupole, we calculate the z-direction Zak phase $\phi_z$ distributions for a quarter BZ in Fig. 4A. As can be seen, the value of $\phi_z$ winds around the straight nodal lines located at the BZ center and hinge. Quantized $\phi_z$ values are found along the diagonal line of the quarter BZ, which is the projection of $C_2T$-invariant plane indicated in Fig. 4B (left panel). The winding of Zak phase around the straight nodal line is a manifestation of the in-plane Berry curvature distribution according to $\int \Omega_\parallel ds = \phi_z^{\theta_2} - \phi_z^{\theta_1}$, as governed by the Stokes' theorem, and illustrated in the right panel of Fig. 4B. We show the full winding of Zak phase around the M – A nodal line in Fig. 4C, where the Zak phase is found to span an approximate range of π and the phase gradient changes sign while crossing each quadrant. This winding range and behavior indicate a nearly π-quantized quadrupole of Berry curvature, where the integral of the Berry flux flowing through a quarter cylinder surface is equal to the Zak phase difference between $\phi_z^2$ and $\phi_z^1$ defined on the mirror planes.

We numerically calculate the Berry curvature distribution in the $k_x$-$k_y$ plane (e.g. $k_z = 0.5\, \pi/c$) and show the results in Fig. 4D. As can be seen, for the 2$^{nd}$ band (shared band of red and blue straight nodal lines), the Berry curvature around both the BZ corners and center has a distribution in the form of quadrupole. Within each quadrant, the Berry flux flows between the two straight nodal lines located at the BZ hinge and center, with directions indicated by the green arrows in Fig. 4D. More interestingly, the strength of Berry curvature quadrupole can be continuously tuned by changing the geometrical parameter, e.g., cylinder height $L_z$ (or equivalently the periodicity in z direction), of the meta-crystal unit structure. For the extreme cases of $L_z$ approaching periodicity $c$ or 0, the inversion symmetry of the system will be restored, which correspond to Zak phase winding of 0 and 2π, respectively. By changing the length of $L_z$ within the limits, the spanned range of Zak phase winding in Fig. 4C can be continuously tuned, as is the strength of Berry curvature quadrupole (detail discussion in Supplementary Material section 8, Fig. S7). This tuning freedom allows for the precise control of Berry curvature in novel phenomena, such as anomalous transport (*55, 56*).

The $D_{2d}$ group allows for both quadrupole and toroidal distributions of in-plane Berry curvature, but toroidal distributions are not allowed by the time reversal symmetry as discussed and verified by the effective Hamiltonian in Supplementary Material section 2, Fig. S1. Dipolar Berry curvature distribution can also be achieved in meta-crystals by breaking the $S_{4z}$ symmetry, as shown for meta-crystals of $C_{2v}$ group in Supplementary Material section 9, Fig. S8. Interestingly, for the case of $C_{2v}$ point group without $C_{2\pm}T$ symmetry, positions of π-quantized Zak phase can still be found, where the presence of 1D π-quantized Zak phase loops is due to a local effective *PT* symmetry (Supplementary Material section 9, Fig. S8).

In conclusion, we proposed and experimentally demonstrated a novel type of quadratic straight nodal lines in the $D_{2d}$ meta-crystal that is protected by roto-*PT* symmetry. These nodal lines are pinned at the BZ center and hinges and serve as sources or sinks of quadrupolar Berry curvature distributions. Truncated systems in all three directions exhibit topological "Fermi-arcs" with super-imaging capability. These "Fermi-arcs" from opposite surfaces intersect at positions of π-quantized Zak phase. The proposed nodal system could find applications in designing novel functioning optical devices or integrated optical systems.

**Acknowledgements:**

This work is supported by the Hong Kong RGC (AoE/P-02/12, 16304717, 16310420) and the Hong Kong Scholars Program (XJ2019007). W.-J.C. is supported by National Natural Science Foundation of China (Grant No. 11874435).

Figures

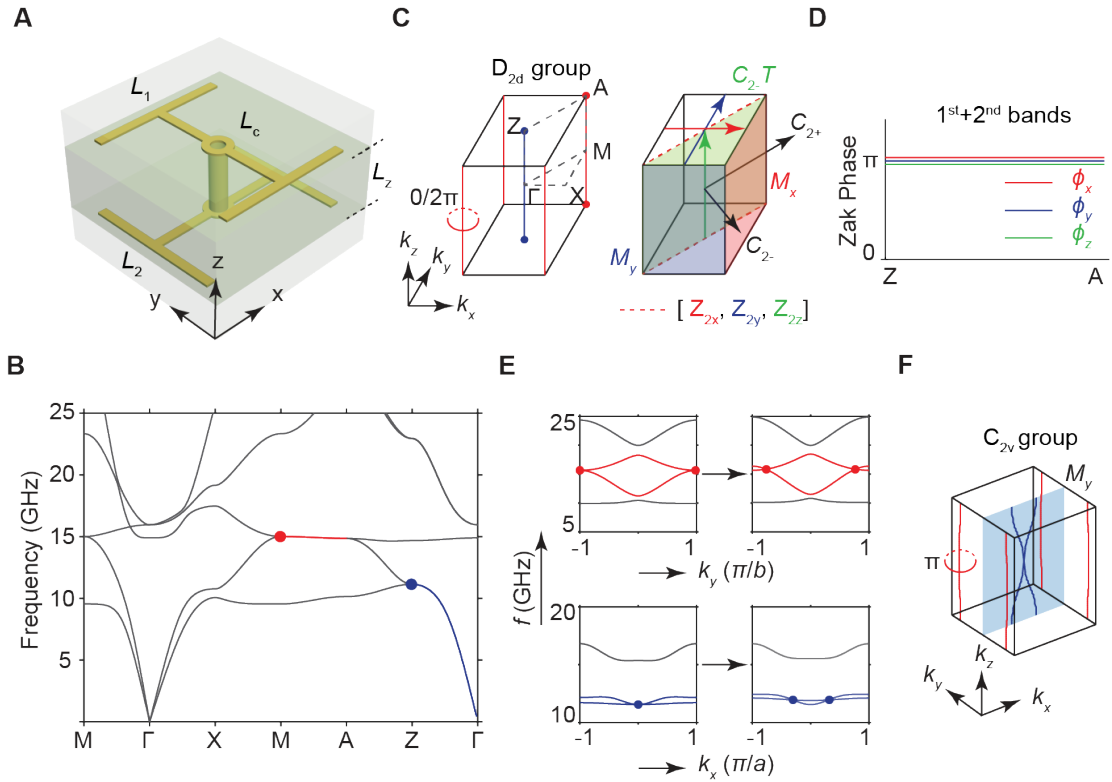

**Fig. 1. Quadratic straight nodal lines protected by roto-*PT* symmetry.** (A) Schematic of a meta-crystal unit cell, the unit cell is repeated along x/y/z directions with periodicity of $a = b = 4$ mm and $c = 3$ mm, respectively. The length of resonator is $L_1 = L_2 = 3$ mm, the length of cross bar is $L_c = 3.5$ mm, and the vertical cylinder is of $L_z = 1.5$ mm height. (B) Calculated band structure for the meta-crystal in (A), degenerate lines can be found along M – A for the 2$^{nd}$ and 3$^{rd}$ bands and Γ – Z for the 1$^{st}$ and 2$^{nd}$ bands. (C) Brillouin zone and straight nodal lines are indicated with red and blue colors, and a 0/2π encircling Berry phase is indicated. In the right panel, the $M_{x,y}$ symmetric planes with quantized Zak phase and the $C_2$-$T$-invariant plane are indicated with red/blue/green colors, respectively. $C_{2\pm}$ axes are also shown. The Zak phase integration paths along $k_x/k_y/k_z$ directions are indicated with colored arrows, and three $Z_2$ invariants can be defined accordingly. (D) Calculated Zak phase for Z – A line along $k_x/k_y/k_z$ directions, which all quantized to be π, thus we have $[Z_{2x}, Z_{2y}, Z_{2z}] = [1, 1, 1]$. (E) Quadratic dispersions are observed for the M/Z points. Breaking $S_{4z}$ symmetry ($L_1 = 2.8$ mm, $L_2 = 3$ mm) splits quadratic line nodes into linear pairs. (F) After breaking $S_{4z}$ symmetry, an in-plane nodal chain (blue) is formed in the $M_y$ symmetric plane, another pair of linear nodal lines (red) is located at the $M_x$ symmetric plane at BZ boundary, the Berry phase of a loop encircling the linear nodal line is π.

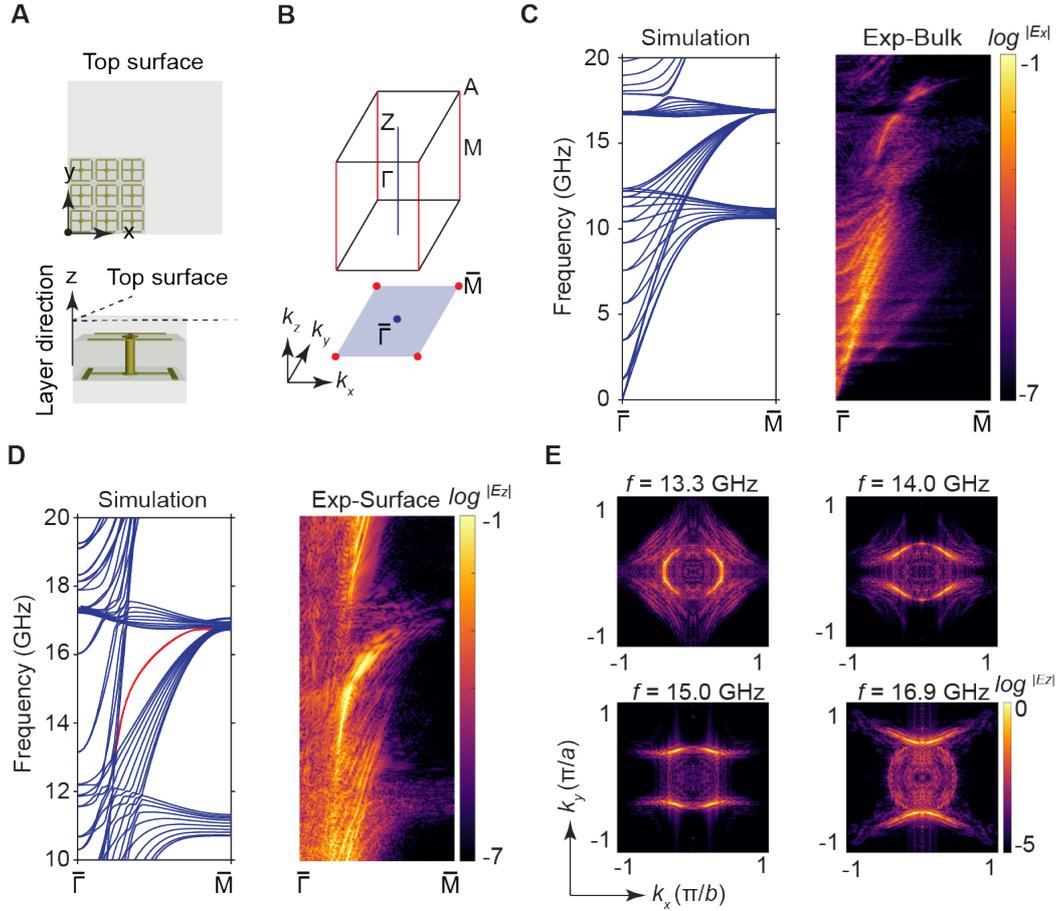

**Fig. 2. Characterizing the straight nodal lines on the (001) surface.** (**A**) Schematic of meta-crystal stacked along the z-direction, and the measured top surface is the x – y plane. (**B**) Surface BZ for the (001) surface, the straight nodal lines project into discrete points. (**C**) The calculated $k_z$-projected band dispersions (left panel), and experimental measurement of band projections with PEC boundary to the right. The straight nodal line at the Brillouin zone center (blue line along Γ – Z in (B)) can be identified with the degeneracies of photonic band pairs at $\bar{\Gamma}$ position. (**D**) Calculated and measured projected bands along $\bar{\Gamma}$ – $\bar{M}$, surface modes are indicated with red color in the left panel. The surface modes connect to the bulk degeneracy at $\bar{M}$. (**E**) Measured EFCs of surface modes on the top surface evolve from closed to open curves.

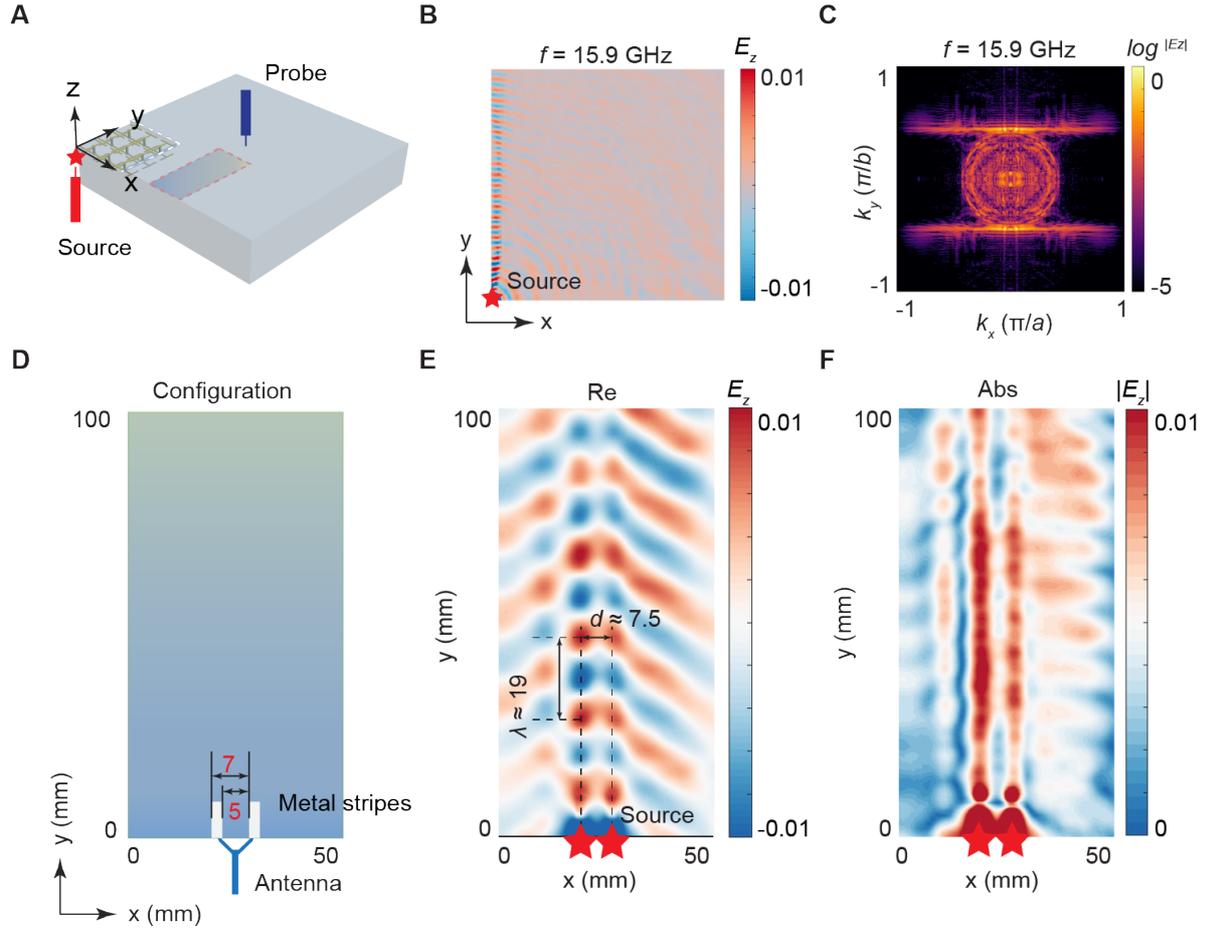

**Fig. 3. Super-imaging with topological "Fermi-arcs".** (**A**) Schematic of meta-crystal arrangement for surface "Fermi-arcs" propagation measurement. Source position is indicated with red star, and microwave antennas are indicated with red/blue colors. (**B**) Field pattern measured with corner source excitation, the excited surface waves are collimated and directed towards the y-direction. (**C**) The EFC of the surface wave at $f = 15.9$ GHz, flat photonic "Fermi-arc" is found. (**D**) Configuration for demonstrating super-imaging. Two metal stripes are allocated on the sample, source antenna is attached to the metal stripes, and they serve as two point sources for surface wave. (**E**) Surface wave propagation at $f = 15.9$ GHz, two collimated beams originating from the metal stripes can be identified. Retrieved propagation parameters are $d \approx 7.5$ mm, $\lambda \approx 19$ mm, imaging capability is demonstrated as $\eta = d / \lambda \approx 0.39$. (**F**) Absolute amplitude of surface wave, and the two beam profiles can be well recognized while propagating.

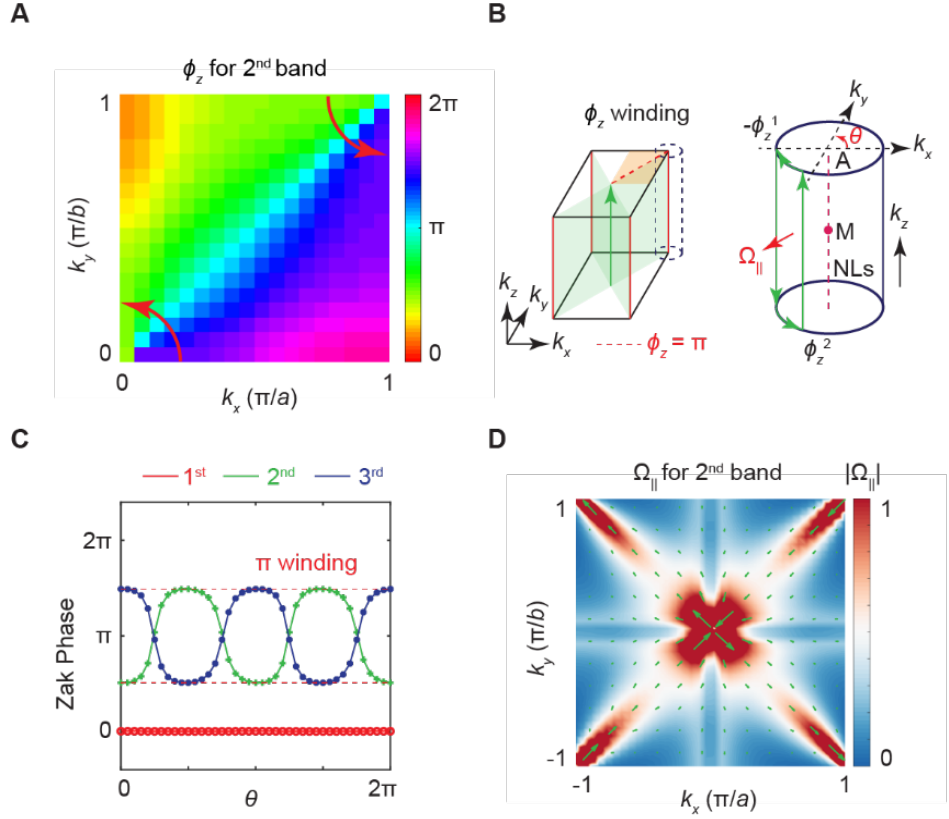

**Fig. 4. Quadrupoles of Berry curvature.** (**A**) Zak phase integrated along the $k_z$-direction ($\phi_z$) for a quadrant of the BZ. The $\phi_z$ Zak phase winding over a quarter circle (red arrows) in $k_x$-$k_y$ plane is approximately $\pi$. (**B**) BZ and hinge nodal lines (red) for the meta-crystal in left panel, $C_{2\pm}T$-invariant planes are indicated with green color, and the arrow indicates the Zak phase integration direction. The winding of Zak phase implies in-plane Berry curvature flux in right panel by Stokes' theorem. (**C**) The Zak phase winding around the nodal line along M – A, the winding range is approximately $\pi$ and the Zak phase gradient changes sign across each quadrant, which indicates a nearly $\pi$-quantized quadrupole of in-plane Berry curvature. (**D**) The in-plane Berry curvature calculated numerically for the 2$^{nd}$ band at the cut-plane of $k_z = 0.5\,\pi/c$. Green arrows indicate the vector directions. Quadrupoles of Berry curvature can be found to be centered at both $\Gamma$ – Z and M – A positions, the Berry curvature flows between them.